# AUTOMATED SPLIT HOPKINSON PRESSURE BAR FOR HIGH THROUGHPUT DYNAMIC EXPERIMENTS

Mouliswar Ramakumaresan, Vladimir Kornev*, Pranav Kartha*, Isaac Faith Nahmad, Suraj Ravindran
[1]Department of Aerospace Engineering and Mechanics
University of Minnesota, Minneapolis, 55455
Corresponding author: sravi@umn.edu (Suraj Ravindran)
* These authors equally contributed to the work

**Abstract**

The design of novel, impact-resistant materials requires extensive datasets to support data-driven design and predictive modeling. While high throughput characterization is feasible at the sample micro and nanoscale, there remains a critical need for an experimental tool capable of facilitating high-throughput measurements at the macroscale. This paper presents the design and development of a fully automated split Hopkinson pressure bar (SHPB) with full-field diagnostics for performing dynamic compression experiments on different materials. The automated SHPB consists of four main components: 1) striker launch and retrieval system, 2) bar repositioning mechanism, 3) automated sample placement system, and 4) diagnostics. Each system integrates electromechanical devices and actuators, enabling fully automated SHPB experiments with minimal human intervention, increasing reproducibility and throughput. A data analysis tool has been developed to automate the post-processing of the data obtained from the setup. The setup also incorporates automated high-speed imaging, enabling full-field strain measurement capabilities. To benchmark the setup, fully automated dynamic compression experiments were conducted on 45 Copper 101 samples, 20 samples made with 1100 aluminum, 20 samples of polycarbonate, and 20 photopolymer resin samples. Stress-strain curves were extracted from the raw data using the automated data analysis tool and validated against conventional SHPB analysis techniques, ensuring both speed and reliability in high-throughput testing. Full-field strain measurements in experiments show strain values comparable to those obtained with strain gages. This automated SHPB system enables dynamic compression experiments at an unprecedented rate of 60 samples per hour, significantly accelerating data generation for high-strain-rate materials research.

**Keywords:** Data-driven, materials design, high-throughput, automation, Hopkinson bar, dynamic compression, high strain rate

## 1. Introduction

Progress in data-driven methods and machine learning is revolutionizing the mechanics of materials research, particularly in constitutive modeling, material characterization, and the design of novel materials [1–4]. These methods require large data sets, which are often limited in traditional experimental methods. Consequently, most of the current data-driven algorithms rely on the data obtained from numerical simulations to develop and validate advanced data-driven methods. While simulations facilitate rapid data generation, they inherently limit the integration of experimentally observed physics, especially in capturing complex deformation and failure mechanisms. To bridge this gap, high-throughput experimentation is essential for generating diverse, experimentally validated datasets to support data-driven materials design.

There have been recent efforts focused on developing high-throughput experiments for characterizing materials across varying strain rates to screen the strength properties of materials for accelerating material discovery [5,6]. Most of these high throughput experimental methods in low and high strain rate regimes are based on well-established nano-indentation and micro-indentation experiments. Indentation techniques typically measure force-displacement curves, which are then analyzed using Hertz's elastic solution to contact problems [7] or advanced inverse analysis for extracting the stress-strain curve of the materials from the force-displacement measurements [8,9]. These nano/micro indentation techniques are fast for screening the yield strength of materials; however, reliably measuring the large strain response of materials through these techniques is challenging. The main drawback of these techniques is that the data analysis is complex and prone to high errors when estimating tensile strength and other material properties beyond the yield strength of the materials [5]. Furthermore, the amount of material deformed in indentation techniques is minimal, which may not represent the continuum-scale behavior of the materials with microstructural heterogeneities at a larger length scale. In high strain rate regimes, indentation techniques have been implemented to study the material strength at strain rates ranging from $10^2$ to $10^4$ /s [10–13]. For strain rates exceeding $10^4$ /s, laser-induced projectile impact testing (LIPIT) has been used to estimate the strength and damage in materials [14–17]. The high-strain-rate nano/micro indentation and LIPIT techniques have several drawbacks, including spatially non-uniform strain rates, complex stress states, and varying strain rates during loading and unloading.

Indentation and LIPIT-based experimental methods are crucial tools for the first-step evaluation of the material yield strength at varying strain rates. However, obtaining highly repeatable data on yield strength, tensile strength, and large-strain behavior of material requires alternative strategies. Automation of traditional experiments like uniaxial tension/compression for static loading and split Hopkinson pressure bar (SHPB) experiments for the high strain rate characterization [18–21]. The primary challenge in these conventional experiments lies in the time-intensive process of loading, aligning, and rearming the apparatus after each experiment. Recently, the uniaxial quasi-static compression experiments have been automated, and can perform 1000s of experiments rapidly [22]. However, the conventional split Hopkinson pressure bar experiments used to characterize material strength at these high strain rates are challenging to automate due to the automation requirements of several processes involved in the experiment.

Several studies have attempted to automate aspects of SHPB experiments, particularly in rearming the striker bar [23]. Efforts have included vacuum-based striker retrieval [24] and flywheel-driven launch and retrieval mechanisms [25]. However, these approaches introduce trade-offs, such as altering the shape of the input stress pulse or limiting experimental flexibility. Furthermore, existing automated setups do not address critical challenges such as sample reloading and bar repositioning, which are essential for accurate and repeatable high-throughput experimentation. Overcoming these limitations requires a fully integrated automation strategy that synchronizes all aspects of the SHPB experimental workflow.

The primary objective of this study is to develop an automated SHPB capable of performing large sets of experiments in the strain rate range of $10^3$ /s to $10^4$ /s. The experimental setup has been designed and manufactured to systematically address the following challenges: (a) automatic re-arming of the striker bar to produce the desired compressive pulse; (b) self-regulated sample placement; (c) automated data acquisition and processing. Section 2 details the experimental setup and the design of individual components, along with a summary of the fully automated data processing technique used to post-process the data. Section 3 discusses the results obtained from experiments on copper, aluminum, polycarbonate, and photopolymer resin samples, followed by the conclusion.

## 2. Materials and Methods

### *2.1 Experimental setup and data analysis*

The split-Hopkinson pressure bar (SHPB) setup consists of three primary components: a striker bar, which generates a compressive stress pulse; an incident bar, which transmits the pulse to the sample; and a transmitter bar for propagating the stress pulse altered by the sample deformation. Under force equilibrium, the stress, strain, and strain rate in the sample are determined using the following equations:

$$\text{Sample stress,}\ \sigma_s(t) = \frac{EA}{A_s}\varepsilon^T(t) \qquad 1$$

$$\text{Sample strain,}\ \varepsilon_s(t) = \frac{-2C_0}{L_s}\int_0^t \varepsilon^R(t)\,dt \qquad 2$$

$$\text{Strain rate,}\ \dot{\varepsilon}_s(t) = \frac{-2C_0}{L_s}\varepsilon^R(t) \qquad 3$$

In the equations, $\varepsilon^T$, $\varepsilon^R$ are the transmitted and reflected strain signals. Parameters $A$ and $A_s$ are the cross-sectional area of the bar and sample; $E$ and $C_0$ are the Young's modulus and elastic wave speed in the bar material. Strain gages are used to measure the strain signals, with two diametrically opposite strain gages on the incident bar and a single strain gage on the transmitter bar. The two diametrically opposite strain gages on the incident bar compensate for bending effects, ensuring accurate measurement of the incident/reflected pulse, and, a single strain gage on the transmitter bar measures the transmitted pulse, as bending is expected to be negligible in this region.

*2.2 Automated experimental split Hopkinson pressure bar*

The fully automated SHPB (ASHPB) system consists of four key components (Fig. 1):

1. The Striker Launch and Retrieval System (SLRS): This system uses sensors and a solenoid-controlled pneumatic system to automate striker launch and retrieval, ensuring consistent impact velocity.
2. The Bar Repositioning System (BRS): This system ensures accurate alignment of the incident and transmitter bars using stepper motor-driven actuators.
3. The Sample Placement System (SPS): This system features a motorized stage to automatically load and position samples between the bars, improving repeatability and reducing setup time.
4. Diagnostics: This system integrates automated strain gage data acquisition and high-speed imaging for full-field strain measurements.

The synchronization of these four subsystems enables fully automated, high-throughput SHPB experiments with minimal human intervention.

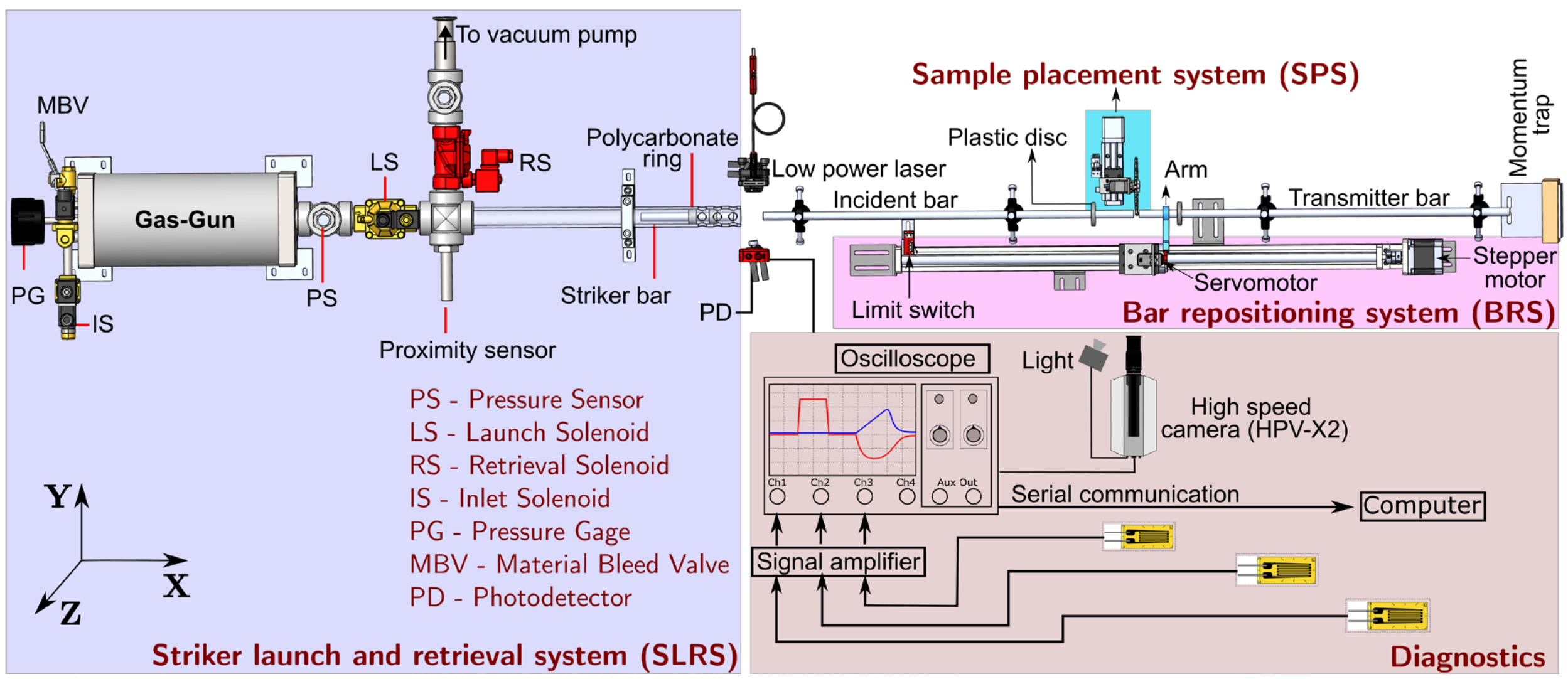

**Fig. 1** Top view of rendered CAD of the complete mechanical system of Automated Split Hopkinson bar with the gas gun system, bar repositioning system, and sample placement system. (components are not to scale)

Four different materials were used for the experiments in this study: copper 101 (Cu101, Xometry), 1100 aluminum (Al-1100, McMaster-Carr product number: 88685K97), polycarbonate (PC), and photopolymer resin samples (Formlabs Grey resin V4). A total of 40 samples were prepared from Cu101, comprising 20 as-received and 20 annealed under an inert argon gas atmosphere at 370°C for 60 minutes. 1100 aluminum sheets and polycarbonate sheets were used to prepare 20 Al-1100 and 20 PC samples, respectively. Vat-polymerization was used to fabricate photopolymer samples with a layer thickness of 100 μm. Samples were designed using SolidWorks 2024 and printed using a Form 3 printer from Formlabs, USA. After printing, all the samples were removed from the build plate and washed in an isopropyl alcohol bath for 2 min, and fan dried for 15 min. Samples were then cured under UV-light for 1 hour and 15 minutes, followed by a gentle polish using 800-grit paper and washed with acetone to obtain a clean surface. All the samples were fabricated to have dimensions of 6 mm × 6 mm × 4 mm, and the dimensional accuracy of each sample, along the gauge length, was ±0.05 mm, with surface flatness maintained at ±0.1 mm on all faces. In experiments with copper, aluminum, and polycarbonate, C350 maraging steel was used as the material for the striker, incident, and transmitter bars. However, for experiments involving 3D-printed samples, aluminum was used for all three bars. This modification was necessary to obtain a measurable transmitted wave signal due to the low impedance of 3D-printed plastic samples. Incident and transmitter bars were fabricated to have a length of 1524 mm for aluminum and 1736 mm for maraging steel. Both types of bars have a diameter of 12.7 mm and are centerless ground to have a uniform cross-section throughout their length. The entire ASHPB system was mounted on a set of optical tables reaching a total length of 5486.4 mm. To mount the bars, a custom machined brass ring was pressure fitted onto an optical ring mount from Edmund Optics (53 mm outer diameter, three screw adjustable, Part number: #03-668), which was then screwed onto the optical table by a post holder and clamp from Newport Corporation (VPH-2-P, PS-F-1.0C-PK).

*2.3 Automated striker launch and retrieval system*

The schematic of the automated gas gun system in the assembly is shown in Fig.1. In contrast to the conventional gas gun system, which only accelerates the striker bar to a specific velocity, the gas gun system developed in this study can also retract the striker bar to its initial position. The primary components of the system are a gas gun, a barrel, and a vacuum pump. The gas chamber is designed to hold gas pressures up to 250 psi, and the barrel has sufficient length for the striker bar to accelerate to the desired velocity. The internal diameter of the barrel is 40 mm, and the striker bar is 12.7 mm. Therefore, to ensure proper fit between the striker and the barrel, a polycarbonate ring with a diameter of 39.95 mm was press-fitted onto the striker; see Fig.1 (in SLRS). These rings were strategically designed to achieve a clearance fit to allow for the negative pressure differential between the open end of the barrel and the vacuum end when the vacuum pump is turned on during retrieval. Pressurization and bleeding of the gas gun, as well as launch and retraction of the striker bar, are regulated by using electromechanical solenoid valves. A high-pressure gas source fills the gas chamber with the help of an inlet solenoid valve (IS) and a pressure sensor (PS) from Amphenol Advanced Sensors (NPI-19J-300G2). A vacuum pump with a two-way retrieval solenoid (RS) retracts the striker bar after each experiment to its initial position by creating negative pressure behind the projectile. While the striker bar travels backward in the barrel, an infrared proximity sensor (Automation Direct: C18D-0P-1E/CD12L-0B-020-A0) detects the arrival of the striker bar. The proximity sensor ensures the position of the striker bar is approximately the same across all experiments, thereby maintaining a consistent striker impact velocity at a given pressure. Once the striker bar is retracted and the required pressure is attained in the gas chamber, the launch solenoid (LS) is actuated to release the compressed gas to launch the striker bar.

*2.4 Bar Repositioning System*

The Bar Repositioning System (BRS) is the drive unit that controls the entire setup to function in synchronization with each other. The BRS unit consists of 2 actuators to provide 2 degrees of freedom, as illustrated in Fig. 1. The first actuator is a linear motorized stage (FUYU Motion FSK40) controlled by a bipolar stepper motor. The travel length of this linear stage is 0.7 meters, and its resolution is 300 micrometers. The second actuator, a DC servomotor, is attached to the linear stage and has an allowed rotational range of 30° about the X-axis. A 3D-printed arm is connected to the servomotor, which can swing in and swing out when required. A polycarbonate ring is pressure fitted onto the incident and transmitter bar to help in the translation of the bars with the 3D-printed arm attached to the servo motor. During the experiments, the bar repositioning system moves the incident bar towards the gas gun to push the striker bar back into the barrel. This push creates a proper seal between the barrel and the projectile, ensuring a vacuum seal for retraction. Then, the incident bar is pushed forward to position the right end of the transmitter bar against the momentum trap, which ensures accurate positioning of the transmitter bar in every experiment. Following this, the incident bar is moved towards the barrel, and the sample placement system moves the samples toward the bars to sandwich the sample between the incident bar and the transmitter bar. Once the sample is sandwiched between the bars, the BRS moves the transmitter bar towards the barrel till the impact end of the incident bar reaches the low power laser. An Atmel Alf and Vegard's Reduced Instruction Set Computer (Atmel AVR) based 8-bit microcontroller controls the entire positioning of the incident and transmitter bar using an open loop control. It is referenced using a limit switch at one end of the linear stage. While the primary purpose of this microcontroller is to control the position of the incident and transmission bars, it also provides a signal to the SPS and, more importantly, acts as the master microcontroller for the gas gun system, ensuring the striker bar fires only after the sample is loaded.

### *2.5 Sample Placement System*

The purpose of this system is to hold and feed samples between the incident and transmitter bars, see Fig. 2. The bar repositioning system controls the sample placement system (SPS). It has been built to have 2 degrees of freedom, one in the Y direction (perpendicular to the bars, see Fig. 1) and the other is the rotation about the X-axis. The SPS consists of a smaller motorized linear stage (FUYU Motion FSL30), actuated by a bipolar stepper motor, which holds another high precision stepper motor (SOYO: SY20STH30-0604A) through an L-clamp as shown in Fig.2. A sample holder is attached to the high-precision stepper motor shaft with slots to load up to 20 samples. The sample holder is 3D printed using a soft material that is compliant and can provide gripping force to samples (6 mm × 6 mm × 4 mm) without damaging the surface finish or interfering with the operation of the setup. Also, the thickness of the holder is smaller than the sample, see top view in Fig.2; therefore, during the sample placement, the sample holder is not constricted by the incident and transmitter bars. The high-precision stepper motor allows for a new sample to be placed between the incident and transmitter bar by rotating 15° to a new sample after every shot. After feeding a new sample, the cartridge is moved away from the incident and transmission bars, using the linear actuator to keep the compliant sample holder from being damaged during the shot. The sample holder is fully loaded with a new set of samples after testing all the previous sample set. Although the sample holder, with a diameter of 85 mm, has been designed to hold up to 20 samples, it is possible to increase this number by either increasing the diameter of the compliant member and/or by reducing the sample size.

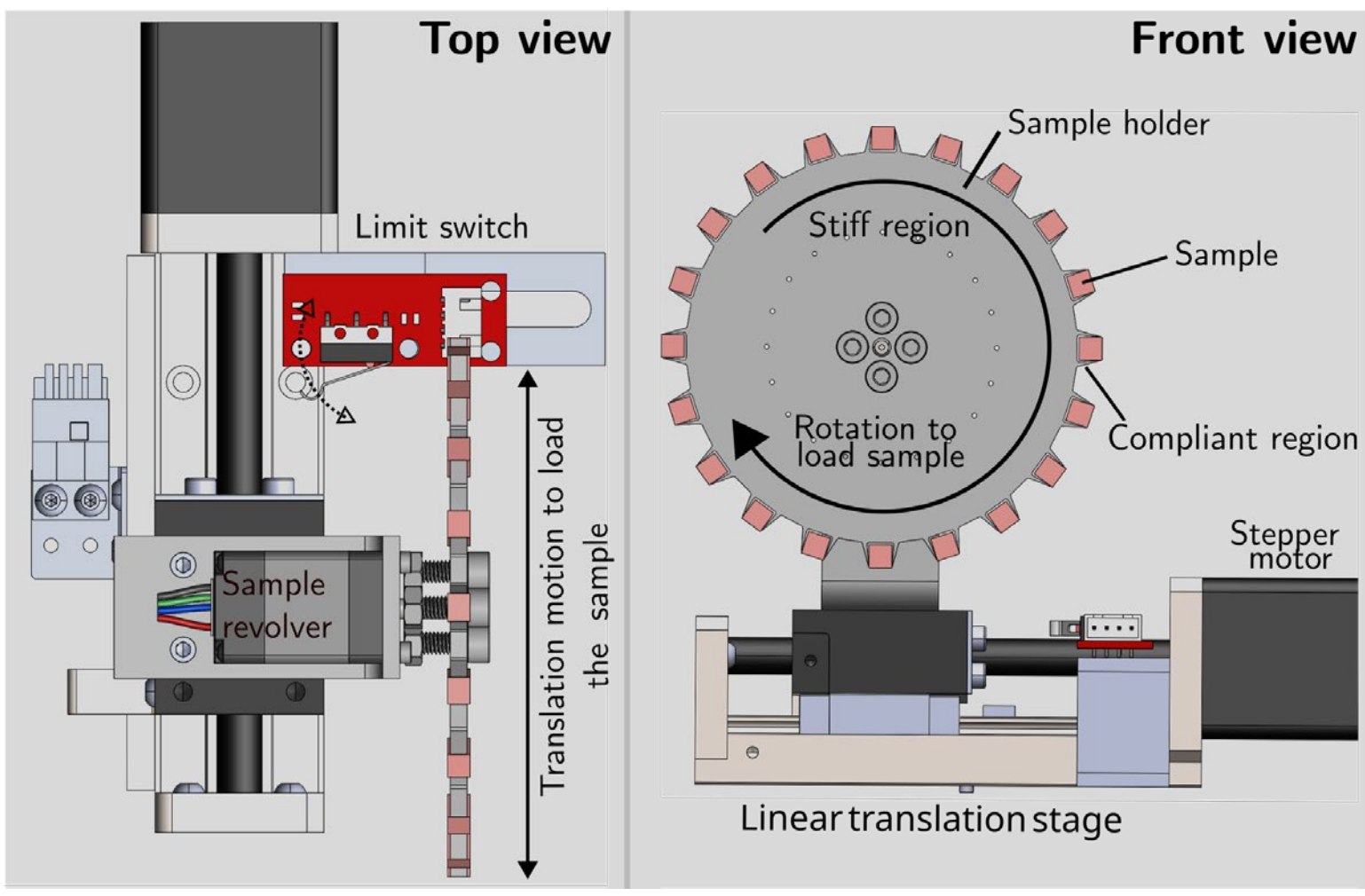

**Fig. 2** Top and front view of the sample placement system.

*2.6 Data acquisition and automated data analysis*

The diagnostics system is an independent unit that consists of a laser source (from Thorlabs, PL202), a photodetector (from Thorlabs, PDA10A2), a 2-series Tektronix Mixed Signal Oscilloscope, strain gages, and a high-speed Shimadzu HPV-X2 camera for *in-situ* digital image correlation (DIC). This unit is designed to perform two main functions: capture signals from strain gages for post-processing and, in addition, provide a trigger signal to the camera to capture DIC data. The entire operation is controlled by a Python script that communicates with the oscilloscope during experiments using serial communication based on SCPI (Standard Commands for Programmable Instruments) commands. The laser source and photodetector provide the first signal to the oscilloscope, triggering data acquisition upon arrival of the striker bar.

A MATLAB script post-processes the data from a series of CSV (Comma-Separated Values) files collected during the experiment, each containing raw strain gage measurements corresponding to an individual experiment. The script iteratively reads and processes each shot, saving results to a Microsoft Excel file while generating consolidated graphs of the true stress-strain response and strain rate evolution for each sample. Processing begins by removing bias from the raw voltage measurements from the strain gages and denoising them using a moving average filter with 10 data points (significantly smaller than the size of strain pulses, ~2000 data points). After that, the voltages from the strain gages are converted to strain using the appropriate conversion factor. Through multiple steps, an algorithm then isolates the incident, transmitted, and reflected signals from the incident and transmitter bars' strain data while also correcting for wave dispersion. This begins by determining the data indices corresponding to the start, first peak, and end of the incident and reflected signals. The start of a signal is found when strain values first pass a signal threshold and then back-track data indices until values reach a zero threshold. The first peak was determined by identifying the point at which strain readings after the signal threshold began to decrease in magnitude. The algorithm would then determine the end of a signal by looking for the point at which the data again reached the zero threshold after the first peak. The signal and zero thresholds are sign-dependent to account for the incident and reflected signals appearing on the same strain gage channel. As a result, the algorithm parses through the incident bar data twice, flipping the sign when looking for the reflected signal.

With the key data indices for the incident and reflected signals, all three signals can be identified and aligned from the incident and transmitter bar strain data. However, this alignment is only preliminary, as wave dispersion stretches the reflected and transmitted signals relative to the incident signal, and dispersion correction shifts the data, requiring realignment. This calls for capturing the signals with a buffer to ensure that no data is lost after dispersion correction and that there are enough data points to properly crop the dispersion corrected signals. This is accomplished by determining the rise time of the incident and reflected signals. The signal with the larger rise time serves as the baseline. This signal is referred to as the reference signal, and its rise time is used as the buffer. To capture the whole pulse with the buffer, the incident and reflected signals are cropped from the incident bar strain data by offsetting a buffer array size worth of data ahead and behind the signal start and end. This incident/reflected data is then time-shifted to align their peaks and array length. The transmitted signal is captured by aligning it with the reflected signal and cropping it to the same size.

Dispersion correction is done on the signals with a buffer using Bancroft's dispersion relationship and the process described in [26]. To characterize Bancroft's dispersion data, the curve fitting parameters corresponding to Poisson's Ratio of the incident and transmitter bars were used [27]. This curve fit describes the dispersion relationship and allows for the determination of the phase velocity for each Fourier component of a pulse, up to the $50^{th}$ component, which is used to find the phase increment for each component. The Fourier expansion of the signals is determined, and the incident signal is propagated forward, while the reflected and transmitted signals are propagated backward using the found phase increments. The signals, now dispersion corrected, are converted back to the time domain and aligned with each other in the same manner as before, with no buffer surrounding the signals. With the fully processed incident, transmitted, and reflected strain signals, quantities of interest (stress, strain, strain rate, etc.) are calculated for each shot and aggregated.

### *2.7 Process synchronization and complete experimental workflow*

The schematics of the process synchronization and experimental workflow are shown in Fig. 3. All the solenoids and the vacuum pump are controlled by DC relays (Songle: SRD-05VDC-SL-C), which are connected to a slave Atmel AVR 8-bit microcontroller. The master microcontroller from the BRS controls this slave microcontroller, enabling synchronous and collaborative operation of BRS, SPS, BLRS, and diagnostics.

The automation begins with the BRS, which translates the incident and transmitter bar to allow the sample placement system to position the sample between them. Both the BRS and SPS rely on limit switches (L1 and L2) to position the actuator stages, ensuring that the incident and transmitter bars return to their preset axial coordinates after each experiment. During the repositioning of the bar with the help of the BRS, the incident bar is moved towards the gun, pushing the striker into the barrel to facilitate its retraction. Following this process, the BRS pushes the incident bar forward, causing it to contact and push the transmitter bar until it reaches the momentum trap at the far-right end. After that, the incident bar is again pushed backward to make the space between the incident and transmitter for the placement of the sample. At this point, the SPS gets the signal from the master controller to move forward and place the sample between the incident and transmitter bars. Once the sample comes between the bars, the BRS moves the incident bars forward, pushing the sample against the transmitter bar with a small force. At this point, the master controller sends a signal to the SPS, which essentially loads the sample between the bars (see the supplementary video for the complete process). Once the sample is loaded and the impact end of the incident bar is placed at the low-power laser, the master controller transmits a signal to the slave controller to take control of the striker launch and retrieval system (SLRS). The vacuum pump attached to the barrel starts operating along with the open-retrieval solenoid (RS), creating a low pressure behind the striker, allowing the striker bar to retract. At one end of the barrel connected to the pressure tank, a pressure sensor (PS) and a proximity sensor (P1) are used to verify the pressure in the gas chamber and the location of the striker bar, respectively. The proximity sensor detects the presence of the striker bar as it is fully retracted. The signal from the proximity sensor shuts the retrieval solenoid valve (RS) and deactivates the vacuum pump. The gas chamber is pressurized to the required level by actuating the gas inlet solenoid valve (IS). Following this, the launch solenoid (LS) is actuated to launch the striker bar. A photodetector is positioned precisely in front of the plane where the striker bar impacts the incident bar. The output change from the photodetector triggers the oscilloscope to record strain gage signals from the incident and transmitter bar, and the Python script saves them automatically to an external storage. The entire process is programmed to run till all the samples are tested. The strain gage data are then imported into the MATLAB script to post-process and generate batch reports. In the experiments with integrated high-speed imaging, the cameras are triggered by the Photodetector (PD, shown in Fig. 2), as discussed in the previous sections. The camera automatically resets after every shot, using the PYAUTOGUI library in Python to capture the images for the following experiment. These images are saved on the computer automatically and later used for post-processing.

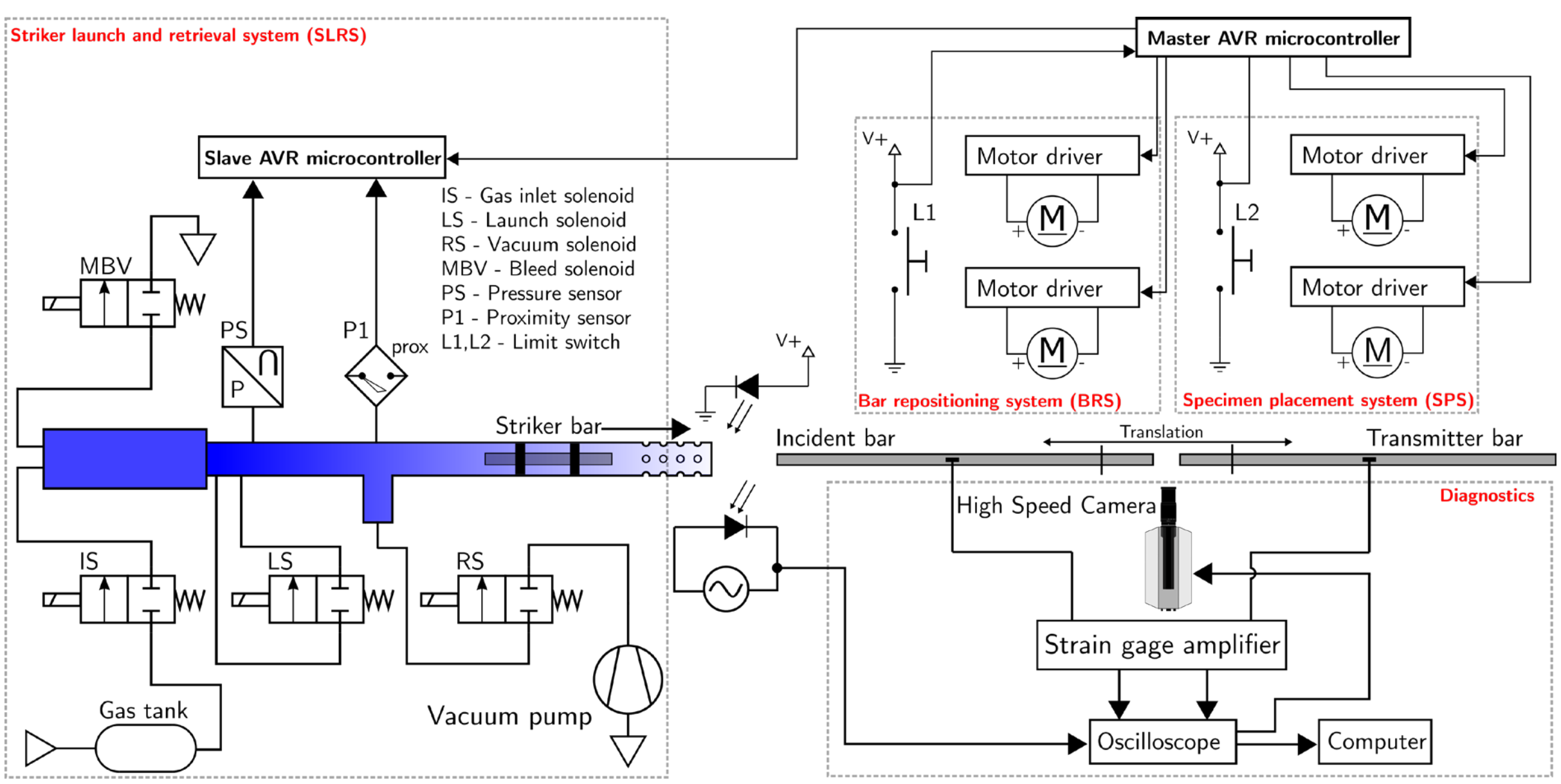

**Fig. 3** Schematic showing the experimental workflow for one batch of samples

## 3. Results and Discussion

To validate and benchmark the experimental setup, five batches of experiments were performed on Cu101, Al-1100, PC, and photopolymer resin samples. The pressure in the gas chamber was programmed to reach 25 psi for Cu101, 15 psi for Al-1100, 15 psi for PC, and 15 psi for photopolymer resin samples during each shot. These pressure values were selected based on the desired strain rate during compression. A pressure gauge (PG, shown in Fig. 2) was used to physically verify the pressure in the chamber for each shot when needed. This was done for two purposes: (1) to verify that the pressure sensor (PS, shown in Fig. 2 and Fig. 3) is calibrated; (2) to map the changes in the pressure for each shot and correlate it with the post-processed data, such as strain rate, true stress, and true strain. The raw signals from the experiment were post-processed using an automated data analysis tool, as discussed in the previous section, to appropriately clip and time shift the incident, reflected, and transmitted pulses for calculating the true stress-true strain and the strain rate in the samples.

### *3.1 Dynamic compression of Cu101 and Al-1100*

Copper has been extensively studied under various strain rate conditions, and in this study, it serves as a benchmark for the Hopkinson bar setup [18,28–33]. Two batches of Cu101 were precisely manufactured to a gauge length of 4 mm and mounted on the sample holder. A successful automated experiment was defined by meeting three key objectives: (a) loading the sample co-axially between the bars; (b) pressurizing the chamber to the programmed value; (c) successfully capturing and recording the incident, reflected, and transmitter bar strain signals for post-processing.

In the first batch of experiments with as-received Cu101, 16 out of 20 experiments were successful, with a total experimental runtime of 19 minutes. Figure 4a shows the strain gage signals from the incident and transmitter bars, which are clipped using the automated data analysis tool developed in this study. Force equilibrium in SHPB experiments is essential for calculating a reliable stress-strain response from the strain gage signals measured in the incident and transmitter bar. The sum of the incident and reflected signals (corresponding to force on the left side of the sample) and the transmitter signals (corresponding to force on the right side of the sample) for all the experiments conducted on the as-received Cu101 are plotted in Fig. 4b. These plots show the attainment of force equilibrium in all experiments conducted on the Cu101 samples. The average strain rates reached in as-received Cu101, were found to be 1051.52 /s, with a standard deviation of 119.61 /s, see Fig. 4c. A summary of the flow strength at different strains and their standard deviation is shown in Table 1. A high variability in the stress-strain response was observed in these experiments, as shown in Fig. 4d, which mainly shows two sets of stress-strain curves with mean peak flow stresses of 450 MPa and 370 MPa at 10 % strain.

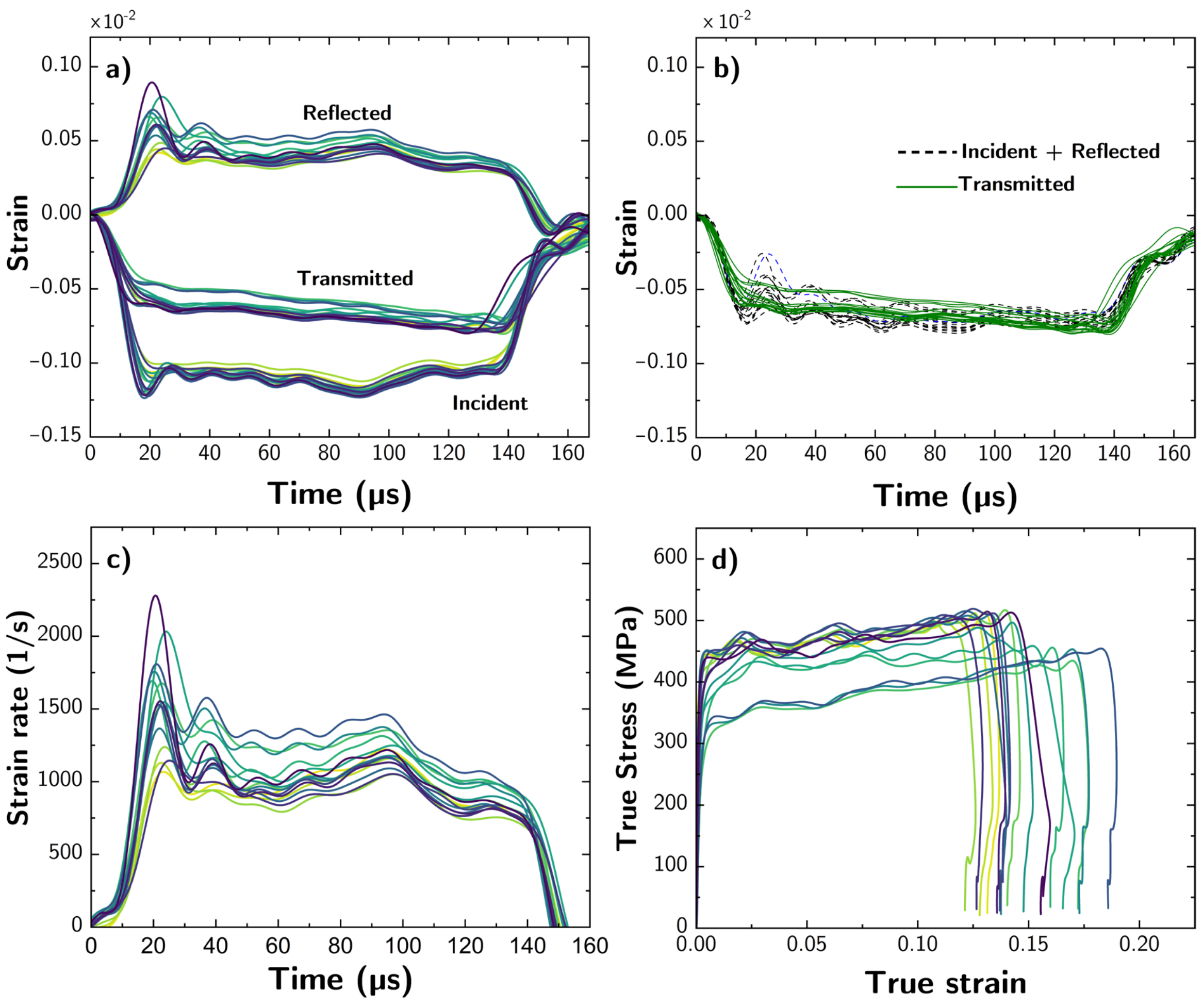

**Fig. 4** Post-processed data from as-received Cu101 (16 samples): (a) dispersion corrected strain gage signals, (b) force equilibrium in the sample, (c) variation of strain rate with time, (d) true stress-strain response

| Parameters | Mean | Standard Deviation |
| --- | --- | --- |
| Average strain rate | 1051.52 /s | 119.61 /s |
| Stress at 2% strain | 396.47 MPa | 74.65 MPa |
| Stress at 5% strain | 429.02 MPa | 37.81 MPa |
| Stress at 10% strain | 453.26 MPa | 34.65 MPa |
| Stress at 15% strain | 439.44 MPa | 24.30 MPa |
| Stress at 20% strain | 412.37 MPa | 28.97 MPa |

**Table 1:** Summary of experiments on as-received Cu101 samples

In a conventional Hopkinson bar, the variability in stress-strain response could be attributed to variations in striker bar velocity, improper sample placement, diagnostic errors in strain gage instrumentation, or inconsistencies in sample preparation. However, since the entire setup has been automated, it is highly unlikely that these variations are due to striker bar velocity as the pressure in the gas chamber reaches the programmed pressure within ±1.5 PSI. Additionally, the pressure in the gas chamber was recorded through the pressure gauge and was found to be within ±1.5 PSI for each shot. Issues with sample placement and diagnostics are also unlikely as the sample placement system and diagnostics calibrate themselves after each shot and have demonstrated high precision (see supplementary data for a video showing sample placement for 20 shots). This leaves the only reasonable explanation for this variation to be due to variation in the sample geometry or its inherent microstructure. As mentioned previously, the dimensions of each sample were measured and found to be within ±0.05 mm tolerance, which cannot account for the significant variation. Therefore, the only possible source of variability in the stress-strain response may be the different processing routes of the as-received Cu101. Cu101 can have different hardness ratings depending on the processing, which can significantly affect the stress-strain response. This hypothesis is tested by performing experiments on annealed Cu101 samples. In the annealed Cu101 experiments, 14 of 20 were successful. The stress-strain response of the annealed sample is shown in Fig. 5c, demonstrating a considerable improvement in consistency between shots for annealed Cu101. The mean strain rate from Fig. 5b, was calculated to be 2010.75 /s with a standard deviation of 63.63 /s. The mean stress at various strains for annealed Cu101, showing the variability between the experiments, is within 3 %. The true stress-strain curve of annealed Cu101 is highly repeatable across all samples. The nominal grain size, after etching and examining under an optical microscope, was calculated to be 22 micrometers, which corresponds with its stress-strain flow curve according to the previous study by Meyer et al. [33]. To understand the efficacy of the developed automated code, a well-established SHPB data processing software from Francis et al. [34] was used to post-process one batch of experiments with a Hopkinson bar analysis code and compared with the corresponding automated MATLAB script results, as shown in Fig.5d. These results confirm that the automated data analysis tool developed in our study reliably calculated the stress-strain response.

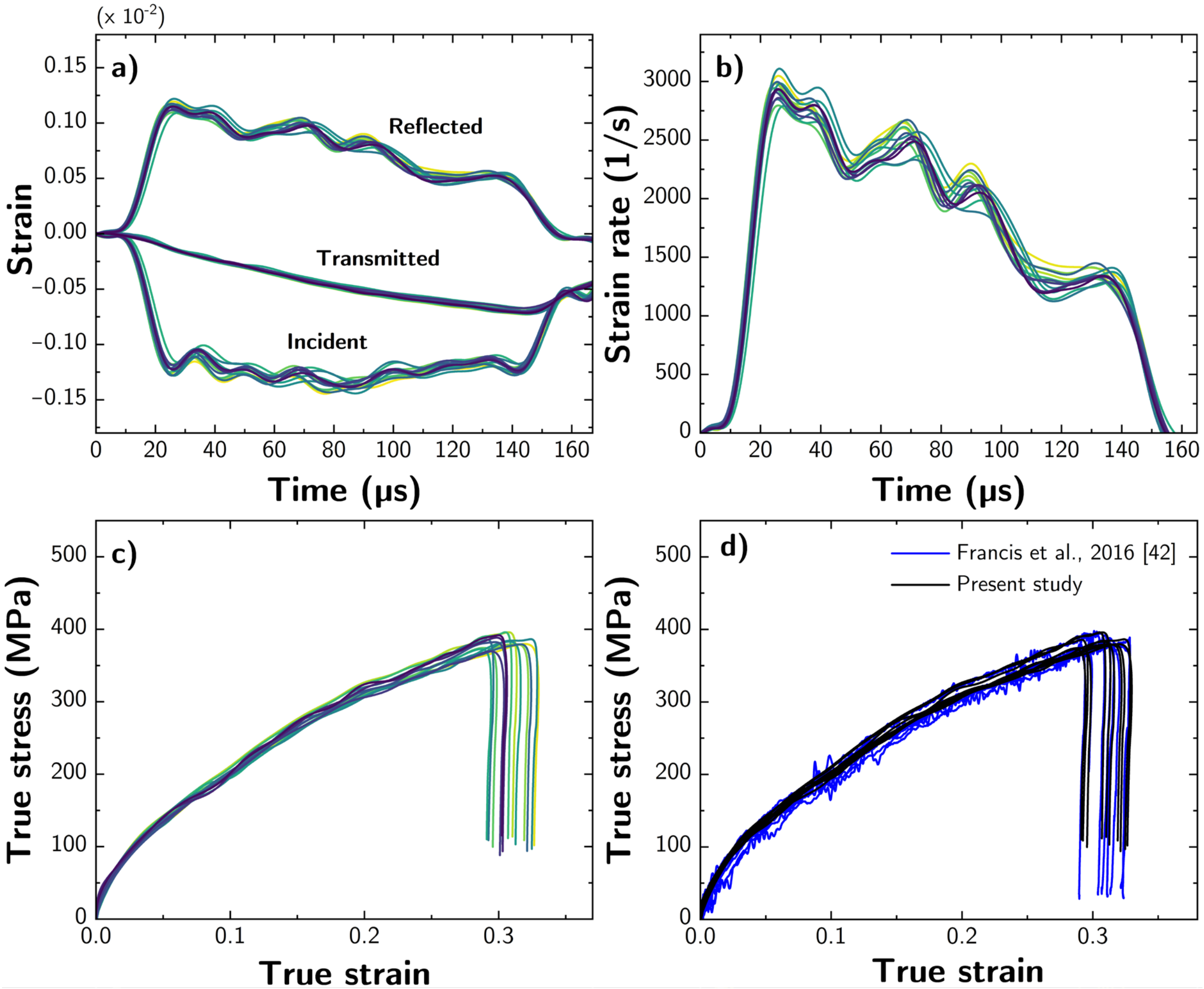

**Fig. 5** Post-processed data from annealed Cu101 (14 samples), (a) dispersion corrected strain gage signals, (b) variation of strain rate with time, (c) true stress-strain response of samples, (d) comparison of true stress-strain curve between Kolsky bar data processing tool (Francis et al.) and automated MATLAB code.

The general trend observed in the true strain rate response from annealed Cu101, in Fig.5b, is due to work hardening, which is commonly observed in elastic-plastic materials. Numerous methods have been proposed to achieve a constant strain rate compression, such as adding a pulse-shaper material before the incident bar, modifying the striker bar to have a tapered cross-section along its length, etc., [21]. The underlying intent is to alter the shape of the trapezoidal incident pulse to a gradually increasing pulse to account for the sample hardening during deformation. While it is straightforward to add pulse shaping material to the automated setup to alter the incident pulse, the objective of the paper is to design and evaluate a fully automated Kolsky bar setup to obtain consistent responses from different materials. To validate this hypothesis, experiments were conducted on Al-1100 samples that do not harden during plastic deformation. Twenty experiments were performed, with 19 successful data collections. The raw data were post-processed using an automated MATLAB script similar to that used for copper experiments. Figure 6 show the results obtained from one batch of Al-1100. Figure 6a and Figure 6b indicate constant strain rate compression of the sample compared to annealed Cu101. Figure 6b shows dynamic equilibrium in all the samples during compression. The average strain rate in Al-1100 was calculated to be 1757.53 /s. The true stress-strain response in Fig.6d indicates a less pronounced strain hardening in Al-1100 compared to annealed Cu101. Table 3 shows the d

Table 2: Summary of experiments on annealed Cu101 samples

| Parameters | Mean | Standard Deviation |
| --- | --- | --- |
| Average strain rate | 2010.75 /s | 63.63 /s |
| Stress at 2% strain | 81.19 MPa | 3.36 MPa |
| Stress at 5% strain | 136.21 MPa | 4.46 MPa |
| Stress at 10% strain | 201.16 MPa | 4.55 MPa |
| Stress at 15% strain | 265.55 MPa | 6.32 MPa |
| Stress at 20% strain | 314.24 MPa | 7.15 MPa |

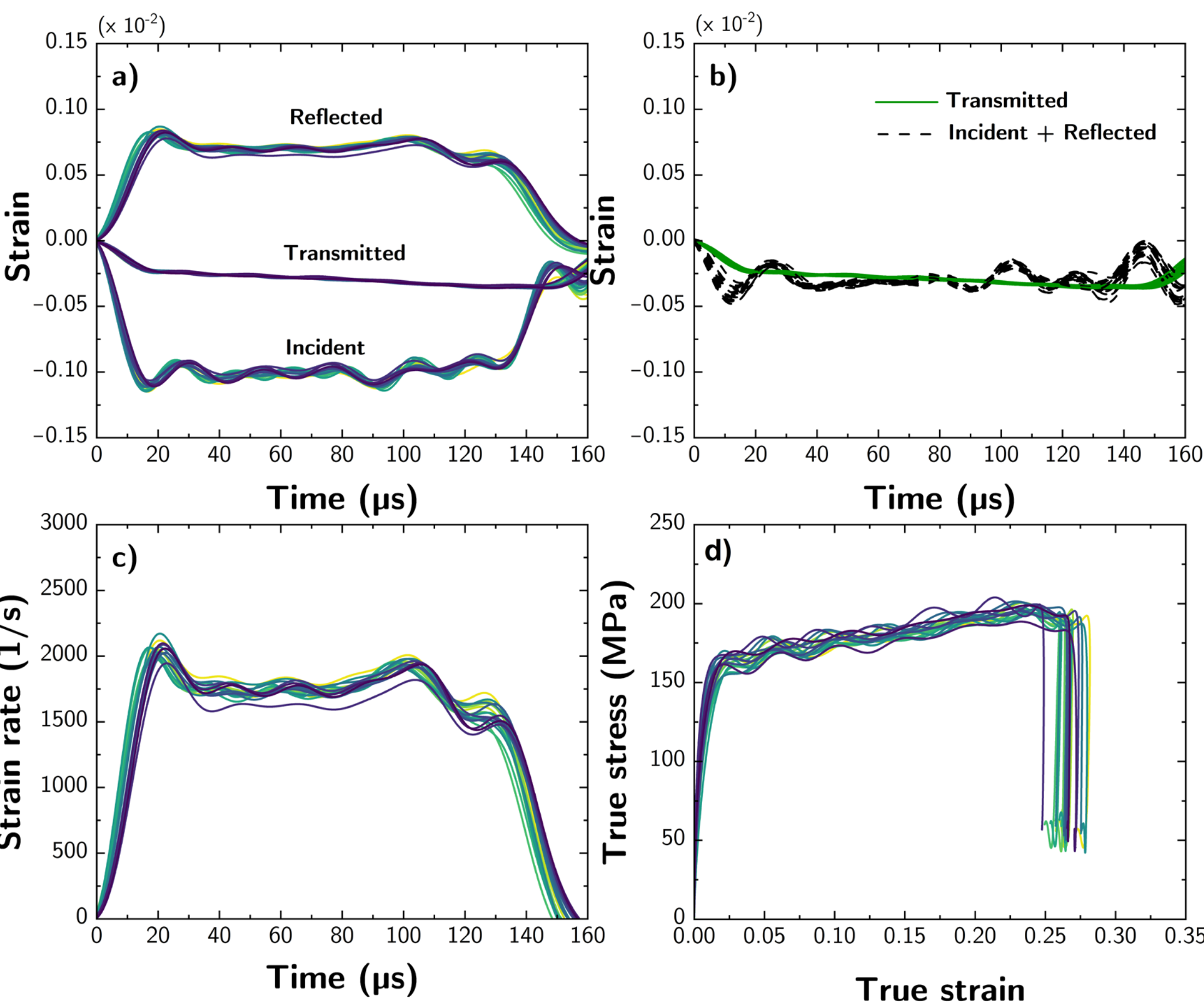

**Fig. 6** Post-processed data from Al-1100 (19 samples), (a) dispersion corrected strain gage signals, (b) force equilibrium in the samples, (c) variation of strain rate with time, (d) true stress-strain response.

**Table 3:** Summary of experiments on Al-1100 samples

| Parameters | Mean | Standard Deviation |
|---|---|---|
| Average strain rate | 1757.53 /s | 41.12 /s |
| Stress at 2% strain | 161.09 MPa | 4.81 MPa |
| Stress at 5% strain | 169.31 MPa | 4.76 MPa |
| Stress at 10% strain | 177.07 MPa | 3.28 MPa |
| Stress at 15% strain | 181.35 MPa | 2.78 MPa |
| Stress at 20% strain | 190.52 MPa | 2.58 MPa |

### *3.2 Automated dynamic compression experiments with full-field measurements on annealed Cu101*

While strain gages are widely used in SHPB experiments, obtaining spatially resolved data is essential for solving many engineering problems. One-dimensional stress-strain data alone is insufficient for developing data-driven material models. New techniques that utilize full-field data to discover material models motivate the need for full-field measurements [35,36]. Therefore, it is necessary to develop automated high speed full-field imaging with DIC capability. To achieve this, a high-speed Shimadzu HPVX-2 camera was integrated into the existing test setup to provide spatially resolved measurements for every experiment on annealed Cu101 through automated 2D DIC measurements (see Fig. 7a). The samples are speckled to facilitate DIC measurements during loading. The images obtained from the experiment were post-processed using commercial software called Vic-2D from Correlated Solutions. Figure 7a shows the full-field strain evolution for three annealed Cu101 samples at similar strain rates. The axial strain is relatively homogeneous for all the experiments, which is an assumption in the one-dimensional analysis of waves in SHPB. Figure 7b shows the average and standard deviation of axial strain for 5 experiments performed using DIC and compared with strain evolution from the strain gage analysis. Both strain measurements correspond with each other, with a slight deviation at about 80 µs. Such observations of material behavior mainly come from the assumptions underlying 1-D wave analysis. These fully automated SHPB experiments with high speed imaging and DIC provide a unique opportunity to generate extensive data from dynamic experiments.

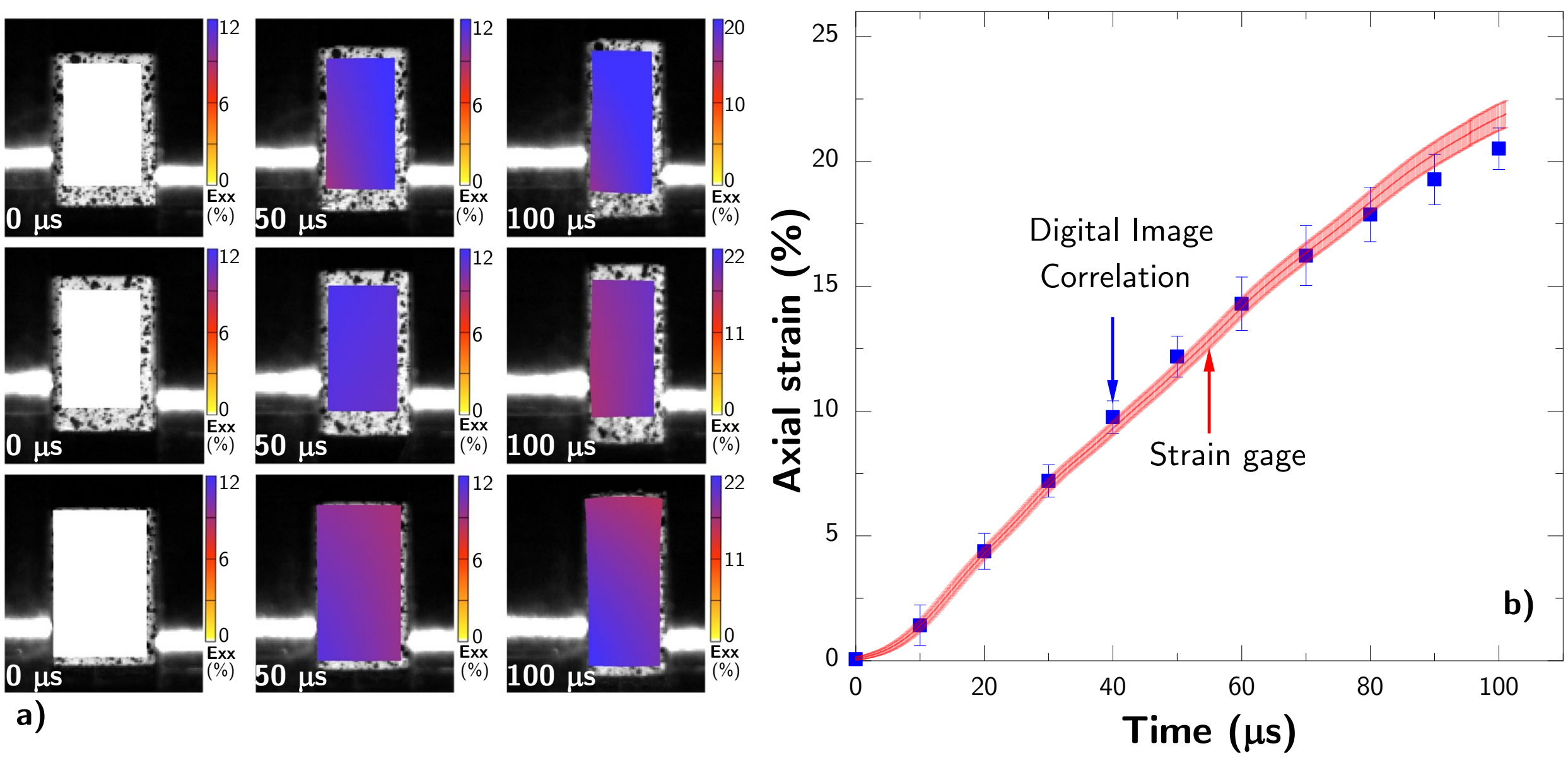

**Fig. 7** (a) Full-field axial strain at three different times, t = 0 µs, 50 µs, and 100 µs, (b) average axial strain evolution from high speed DIC and strain gages in the incident and transmitter bars.

### *3.3 Dynamic compression of photopolymer resin and polycarbonate*

In order show the capability of the experimental setup in characterizing low impedence materials, experiments were performed on 3-D printed and PC samples. Figure 8a shows the variation of incident, reflected, and transmitted strain gage signals plotted as a function of time for the photopolymer resin samples. The plot between strain rate and time, in Fig. 8a, shows an average strain rate of 1829.94 /s with a standard deviation of 160.75 /s. The associated true stress versus true strain response of the samples is shown in Fig. 8b. The mean true stress observed at a strain of 5% is 152.21 MPa with a standard deviation of 8.55 MPa. The general trend, as observed in Fig. 8a, and Fig. 8b, shows scattered signals from the samples used in a single batch. Table 4 provides a quantitative analysis of the experimental results. The variability in the stress-strain response of the samples can likely be attributed to the differences in the curing conditions in each sample. Curing conditions in photopolymer resin fabricated by vat-polymerization influence the mechanical properties of the sample to a large extent [37]. Variation in factors such as intensity of UV-light, exposure time, ambient temperature, print direction, and print defects that arise during cleaning need to be considered and standardized before further experiments can be performed on samples fabricated using vat-polymerization. To validate this, an additional batch of experiments was performed on a PC. Figure 8c and Figure 8d show improved consistency between the samples, and the standard deviation in strain rate was calculated to be 56.96/s/s, as summarized in Table 4.

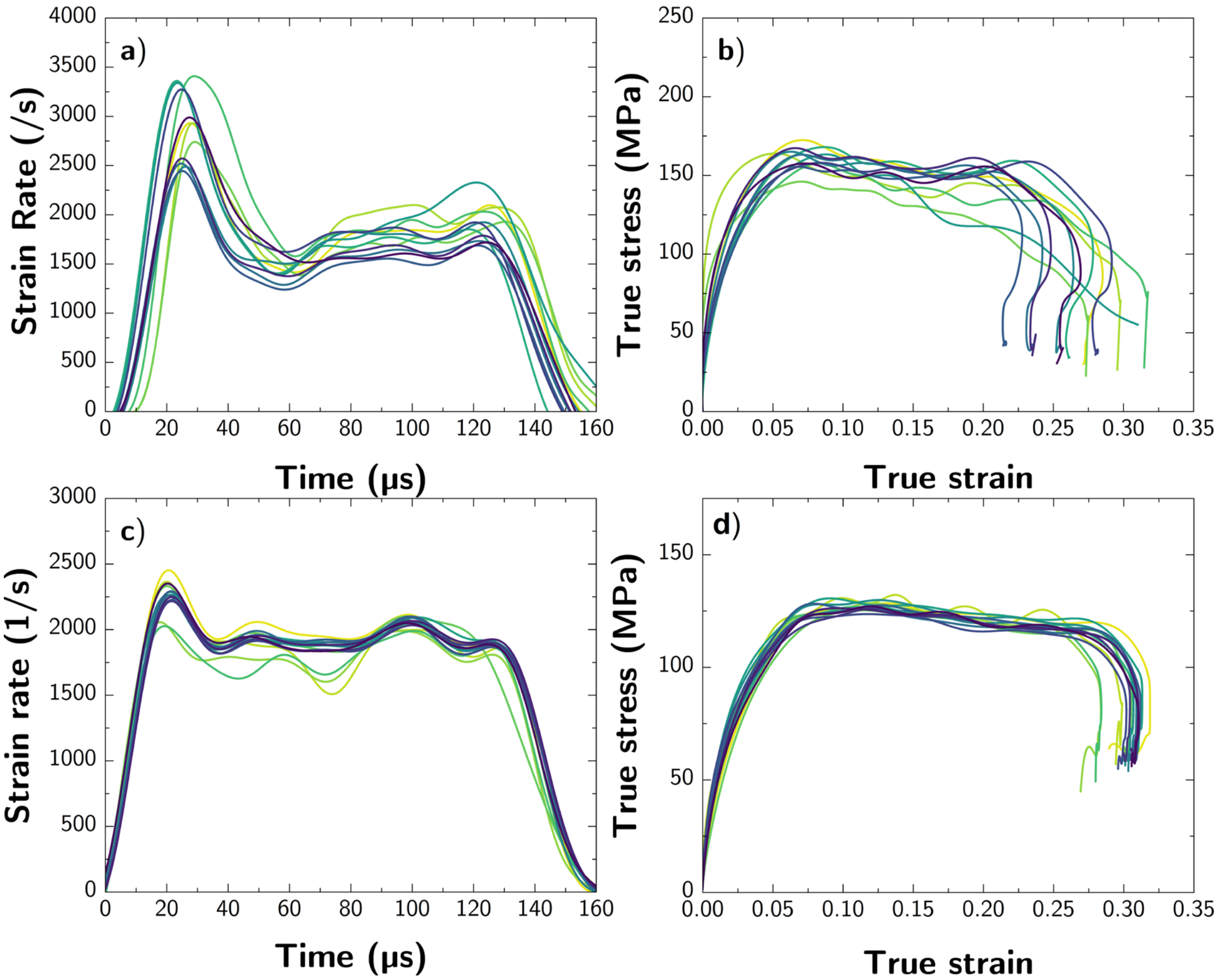

**Fig. 8** Post-processed data from photopolymer resin samples (12 samples), (a) variation of strain rate with time, (b) true stress-strain response; post-processed data from PC (15 samples), (c) variation of strain rate with time, (d) true stress-strain response.

**Table 4:** Summary of experiments on 3D printed resin and PC samples

| Parameters | Mean (photopolymer resin) | Standard Deviation (photopolymer resin) | Mean (PC) | Standard Deviation (PC) |
|---|---|---|---|---|
| Average strain rate | 1829.94 | 160.75 /s | 1927.94 /s | 56.96 /s |
| Stress at 2% strain | 113.23 MPa | 14.33 MPa | 72.89 MPa | 5.21 MPa |
| Stress at 5% strain | 152.21 MPa | 8.55 MPa | 109.82 MPa | 3.84 MPa |
| Stress at 10% strain | 155.55 MPa | 6.75 MPa | 125.92 MPa | 2.16 MPa |
| Stress at 15% strain | 147.31 MPa | 7.81 MPa | 124.54 MPa | 1.89 MPa |
| Stress at 20% strain | 144.64 MPa | 14.08 MPa | 120.29 MPa | 2.29 MPa |

## 4. Conclusion

We have presented a fully automated split Hopkinson pressure bar (SHPB) setup with high-speed full-field imaging to enable high-throughput dynamic compression experiments. This system can perform up to 60 experiments per hour, significantly accelerating data collection for high-strain-rate material characterization. A fully automated post-processing of raw data was implemented using a MATLAB script, and the results demonstrated strong agreement with existing SHPB data analysis tools. Benchmark experiments on Cu101, Al-1100, and PC showed high repeatability, while experiments on photopolymer resin samples showed higher variability in the response. Automated high-speed imaging of annealed Cu101 was used to obtain full-field strain information from DIC, which shows excellent correlation with strain gage data.

Modifications to the setup by using high-accuracy, fast-response pressure transducers in the SLRS will help achieve improved repeatability in striker velocity. The addition of a force sensor to the 3D printed arm in the BRS can provide force feedback to the system, enabling closed-loop control to adjust the force applied to soft samples. With these advancements, the automated SHPB setup could serve as a powerful tool for high-throughput material discovery and characterization. Future work will focus on expanding the range of tested materials, optimizing strain field measurements using DIC, and improving data automation for real-time processing. Additionally, further efforts are needed to refine strain rate control, particularly for soft and polymeric materials, to enhance the consistency of stress-strain data.

In addition to validating the robustness of the system, the experiments also underscore the importance of the presence of a streamlined, automated additive manufacturing process in developing novel materials and structures. Integration of an appropriate additive manufacturing technique into the current automated experimentation setup would significantly expand the applicability, enabling the use of global optimization problems in novel structures where high fidelity data is required for efficient pattern identification. Moreover, incorporating technologies like digital twin, cloud computing, and virtual reality could enhance the scalability of the automated setup, leading to collaboration among researchers

This experimental capability offers transformative potential for the advancement of data-driven material modeling, particularly in regimes where the validation data remain limited. The high throughput experimental setup facilitates the systematic acquisition of extensive datasets characterizing material response under extreme loading rates, thereby substantially reducing the cost and time associated with conventional data collection strategies. Also, the statistically robust datasets generated through this setup enable rigorous model calibration and uncertainty quantification, ultimately enhancing the predictive capability and generalizability of material constitutive models for materials under extreme conditions.

.

**CRediT authorship contribution statement**

Mouliswar Ramakumaresan: Writing – original draft, development of the experimental platform, validation, software, methodology. Vladimir Kornev: development of the experimental platform, validation, review & editing; Pranav Kartha: development of the experimental platform, software, validation, manuscript review & editing; Isaac Faith Nahmad: software, review & editing.
Suraj Ravindran: Writing – review & editing, supervision, resources, conceptualization.

**Acknowledgments**

The research was sponsored by the Army Research Laboratory and was accomplished under Cooperative Agreement Number W911NF2520017. The views and conclusions contained in this document are those of the authors and should not be interpreted as representing the official policies, either expressed or implied, of the Army Research Laboratory or the U.S. Government. Also, the funding support from the University of Minnesota is greatly acknowledged.

**Conflict of Interest**

We have no conflict of interest to disclose.

**Research Data**

Data will be made available on request.